\documentclass[twocolumn,secnumarabic,amssymb, nobibnotes, aps,notitlepage]{revtex4-1}

\usepackage{graphicx}
\usepackage{tabularx}
\usepackage{array}
\newcolumntype{P}[1]{>{\centering\arraybackslash}p{#1}}
\usepackage{multirow} 
\usepackage{makecell} % to allow line breaks in the tables and similar formatting
\usepackage[inline]{enumitem} %inline lists and such
\usepackage{verbatimbox} % allows adding space around tables and such via \addvbuffer command
\usepackage{hyperref}
\usepackage{slashed}
\usepackage{amsmath}

\begin{document}

\title{Imaging baryon number density within the proton}

\author{Spencer R. Klein}
\affiliation{Nuclear Science Division, Lawrence Berkeley National Laboratory, 1 Cyclotron Road, Berkeley, CA 94720 USA}
\email[]{srklein@lbl.gov}

\author{Mathias C. Labonté}
\affiliation{Department of Physics and Astronomy, University of California, Davis, CA 95616 USA}

\author{Gerald A. Miller}
\affiliation{Department of Physics, University of Washington, Seattle, WA 98195 USA}

\author{Zachary Sweger}
\affiliation{Department of Physics and Astronomy, University of California, Davis, CA 95616 USA}

\author{Ramona Vogt}
\affiliation{Nuclear and Chemical Sciences Division, Lawrence Livermore National Laboratory, 7000 East Ave., Livermore, CA 94551}
\affiliation{Department of Physics and Astronomy, University of California, Davis, CA 95616 USA}

\date{\today}%
\begin{abstract}

\end{abstract}

\begin{abstract} 

%{\bf Science version}
%The spatial extent of the proton is usually characterized by its mass and charge radii. However, the spatial distribution of baryon number, a fundamental quantum number of the proton, is unknown and is required for a complete description of the nucleon. We investigate the baryon number distribution in the proton by studying backward ($u$-channel) exclusive meson production arising from photon-proton collisions. Fourier-Bessel transforms are used to analyze the cross sections of backward meson production to find that baryon number is confined to a transverse radius of $0.33 - 0.53$~fm, which is significantly smaller than the known charge and mass radii. Baryon number is concentrated in the center of the proton. 

The spatial extent of the proton is a key factor in nuclear physics.  Different measurement techniques probe different aspects of the proton, yielding different radii. The mass and charge radii depend on the parton and quark distributions respectively, while the mechanical radius depends on the mass/energy distribution.   Here, we probe the spatial distribution of a new proton characteristic, studying the distribution of baryon number within the proton.  
We investigate the baryon number distribution by studying four exclusive meson production channels arising from photon-proton collisions ($\gamma p \rightarrow p \rho^0$, $\gamma p \rightarrow p \omega$, $\gamma p \rightarrow n \pi^+$, and $\gamma p \rightarrow p \pi^0$). The two-dimensional transverse sizes of the interacting systems are extracted  by analyzing the transverse momentum, $p_T$, dependence of the meson production cross section, using Fourier-Bessel transformations. We find that baryon number is confined to a transverse radius of $0.33 - 0.53$~fm.  In comparison, the transverse radius of the proton charge and mass distributions are considerably larger, at least 0.67~fm.  The baryon number is concentrated in the center of the proton.

\end{abstract}
\maketitle
The proton size depends on the particular characteristic  to be measured.  Its charge, mass and mechanical radii are sensitive to different properties of the proton.  However, the proton has other intrinsic characteristics such as baryon number, a conserved quantity. Although other measurements have detailed the size of the proton mass \cite{Duran:2022xag}, charge \cite{Gao:2021sml}, and mechanical radii \cite{Burkert:2018bqq}, the distribution of baryon number remains experimentally unconstrained. Specifically, it is an open question whether baryon number is uniformly distributed within the proton or concentrated in its center.

In Quantum-Chromodynamics (QCD)  the baryon number of a proton is carried by the three net valence quarks. In the parton model, baryon number $B$ can be defined by the net quark number, $B = (1/3)\int_0^1 dx (q(x) - \overline q(x))$, where $q(x)$ and $\overline q(x)$ are the densities of quarks and antiquarks respectively, carrying momentum fraction $x$.  However, the story may be more complex. %because hadrons are made of quarks and gluons and 
Because QCD is constructed from the principle of local color gauge invariance, gluons cannot be ignored when considering interacting hadrons. 

In 1996, Kharzeev \cite{Kharzeev:1996sq} employed the requirement of local gauge invariance of the proton wave function, implying that the baryon is represented by string operators acting on the quark fields, to argue that there should be a substantial baryon asymmetry in the central rapidity region of ultra-relativistic nucleus-nucleus collisions.   The %influence  of the 
baryon number could be connected  with a nonperturbative configuration of gluon fields localized at the junction of these strings~\cite{Rossi:1977cy}. 
%In this picture, 
Then, the baryon junction, the carrier of baryon number, is a configuration of low momentum gluons closely surrounded by  quarks. 

Baryon stopping \cite{Busza:1983rj} in heavy-ion collisions can be described using
%in terms of 
the baryon junction model which makes specific predictions about the rapidity distribution of net baryon stopping \cite{Vance:1997th,Vance:1998vh}.  Data on baryon stopping support the baryon junction picture %, see Ref.
~\cite{magdy2025}. 
%and references therein for a recent overview.  
For example, 
STAR  analyses of baryon stopping in  $^{96}_{44}{\rm Ru} + ^{96}_{44}{\rm Ru}$ and $^{96}_{40}{\rm Zr} + ^{96}_{40}{\rm Zr}$ collisions %show 
are in qualitative agreement with baryon junction models but disagree with models where the valence quarks carry charge and baryon number \cite{tracking}.

Here, we analyze data from a very different type of reaction, exclusive backward meson production \cite{Berger:1971zz,Gayoso:2021rzj}, which can also probe the localization of baryon number.  These reactions are characterized by a large, near-maximal, transfer of momentum from the incident proton to the struck baryon.   This momentum transfer is distinct from conventional forward exclusive photoproduction where the transfer is small. As with baryon stopping, %in 
backward production shifts the baryon %is shifted 
by many units of rapidity  \cite{Cebra:2022avc}. Forward and backward production may be linked in `dual' models \cite{Storrow:1983ct}. 

\begin{figure*}[!t]
\begin{center}
\includegraphics[width=0.9\textwidth]{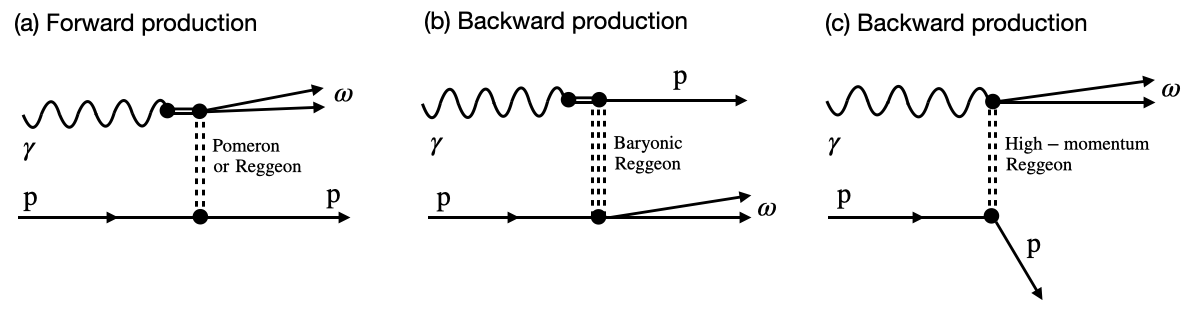}
\end{center}
\vskip -0.1 in
\caption{Diagrams for (a) forward production and (b, c) two possible processes for backward production.  In forward production, (a), $|t|$ is small while $|u|$ is maximized.  Backward production can occur via one of two processes, as shown in (b) and (c).  In (b), production occurs via the exchange of a Reggeon that carries baryon number but small momentum while in (c), the Reggeon carries no baryon number but large momentum.}
\label{fig:diagrams}
\end{figure*}
Exclusive photoproduction is often described in terms of Regge exchange, where the reaction proceeds via the exchange of a pseudoparticle that carries the required quantum numbers such as spin and charge.
Figure~\ref{fig:diagrams} compares Reggeon-exchange models for backward production with forward production.       Figure~\ref{fig:diagrams}(a) shows conventional forward production: a photon interacts with a Reggeon or Pomeron from a target nucleus and forms a vector meson.  Because the Pomeron or Reggeon carries small momentum, Mandelstam $t$ is also small.  Backward production can occur via processes such as those shown in  Fig.~\ref{fig:diagrams}(b) or (c).   Figure~\ref{fig:diagrams}(b) shows the exchange of a Reggeon that also carries baryon number.  %Here, the 
This Reggeon also has small momentum but now the Mandelstam $u$ is small while $t$ is large,  near %its 
maximum.  In Fig.~\ref{fig:diagrams}(c), the exchanged Reggeon does not carry baryon number.  Instead, it carries a very large momentum which transfers most of the incident proton momentum to the produced meson. 
This high momentum transfer implies that, longitudinally, the target is very small, {\it i.e.}\ 
requiring either all three valence quarks in close proximity, or a very high $x$ parton target that subsequently transfers significant momentum to all three valence quarks.  However, this picture does not explain why the momentum transfer is peaked at small $|u|$. This, along with the suppression expected for reactions involving large momentum transfer, makes a process like that %shown 
in Fig.~\ref{fig:diagrams}(c) unlikely.

An alternate approach to these reactions uses the formalism 
of Generalized Parton Distributions (GPDs) ~\cite{Diehl:2003ny}. The lower vertices of Figs.~\ref{fig:diagrams}(a) and \ref{fig:diagrams}(c), involving initial and final state protons, resemble the matrix elements that appear in GPDs. Here Transition Distribution Amplitudes (TDAs) \cite{Gayoso:2021rzj, Lansberg:2011aa,Pire:2021hbl,Pire:2025wbf} can be used to model meson production. 
The TDA, a matrix element of a three-quark color-singlet operator between a nucleon and a meson, can potentially describe the lower vertex of Fig.~\ref{fig:diagrams}(b).  Thus TDAs naturally enter into expressions for backward production scattering amplitudes. 

Backward photoproduction has been studied for several meson final states.  Unfortunately, many of these data were taken at low photon energies, {\it i.e.}\ in the region where baryon resonances can dominate production; over a narrow range of $u$; or with small statistical samples, all of which can complicate interpretation of the results.  We focus on four different reactions on proton targets: backward production of $\pi^0$, $\pi^+$ (leaving a neutron in the final state), $\rho^0$ and $\omega$.   Although electroproduction data exist at different values of $Q^2$ (photon virtuality) there are insufficient data for a systematic comparison.  Therefore, we focus on photoproduction here.   The Electron-Ion Collider should fill the need for electroproduction data, with systematic measurements of backward production of multiple mesons over a wide range of energy and $Q^2$ \cite{Cebra:2022avc,AbdulKhalek:2021gbh,Sweger:2023bmx}.

As Fig.~\ref{fig:diagrams} shows, backward production (Figs.~\ref{fig:diagrams}(b) and (c)) are similar to forward production of the same meson (Fig.~\ref{fig:diagrams}(a)).  The kinematics are similar, albeit with Mandelstam $u$ and $t$ interchanged, where
 %$t$ and $u$ are
$t = (p_1 - p_3)^2$ and $u = (p_1 - p_4)^2$.  The four-momentum $p_1$ refers to an incoming particle, $p_2$ is the target proton, and $p_3$ and $p_4$ refer to outgoing particles, the produced vector meson for $t$ ($p_3 \equiv p_V$) and the outgoing proton for $u$ ($p_4 \equiv p_N$), see Fig.~\ref{fig:diagrams}. 
%Several different  models have been used to address backward production, see Ref.~\cite{Gayoso:2021rzj}.  %Many of these models are also relevant for baryon stopping in more complex collisions.

In Regge exchange models, the $\gamma N$ cross section may be parameterized following \cite{Sweger:2023bmx,Cebra:2022avc},
\begin{equation}
\frac{d\sigma(Q^2,W,X)}{dX} =
\frac{A\Lambda^8\exp{[-D(X-X_{\rm min})}]}
{(W^2-m_p^2)^2(Q^2+\Lambda^2)^4
%/{\rm GeV^4}
}
\label{eq:regge}
\end{equation}
where $X$ is $u$ or $t$, depending on the preferred reaction; $W$ is the photon-nucleon center of mass energy; $m_p$ is the proton mass; 
%$Q^2$ is the squared momentum transfer of the photon; 
$X_{\rm min}$ is the kinematic limit for a given $W$ and final state meson mass; and $\Lambda^2$, $A$ and $D$ are free parameters, determined by fits to data.  Usually, $\Lambda^2 = 2.77$ GeV$^2$, appropriate for a dipole form factor \cite{Sweger:2023bmx,JeffersonLabHallA:2008tza}. 
$\Lambda^2$ is an independent parameter only for electroproduction via virtual photons.  Here $A$ normalizes the overall production rate while $D$ is related to the size of the target.  The simplified exponential dependence on $X$ works well for proton targets but can fail for ion targets.  It is also likely to fail at large values of $|X|$, where reactions probe the internal partonic structure of the nucleon rather than its density distribution.

Equation~(\ref{eq:regge}) can hold for either $d\sigma/du$ for backward production or $d\sigma/dt$ for forward production with $X_{\rm min} =u_{\rm min}$ or $t_{\rm min}$.  In both cases, in the high energy limit, the energies of the incoming and outgoing particles are approximately equal so that $t$ and $u$ are equivalent to the squared transverse momentum, $t \sim u \sim -p_T^2$.  However, in the data used here, because $W < 6$~GeV, an exact relationship between $t$ or $u$ and $p_T^2$ is needed.  It is discussed in the supplemental material.

The 
%Regge trajectory parameterized 
parameterization in Eq.~(\ref{eq:regge}) 
uses only a few parameters.  However, the exponential $X$ dependence misses the diffractive minima observed in some datasets.  It also provides relatively little insight into the underlying meson production mechanism.

Different reactions may occur via different Regge trajectories, which are associated with the exchange of different pseudoparticles carrying quantum numbers.  In the case of high energy forward production, the Pomeron, which carries the quantum numbers of the vacuum, dominates.  To lowest order, the Pomeron is assumed to be composed of two gluons.  At higher orders, quark contributions may be introduced.  In a quark-based model, light baryon Regge trajectories may be thought of as involving diquark exchange \cite{Simonov:1989ff}.  Diquarks are colored two-quark states connected by an attractive potential.  However, diquarks cannot explain
reactions where the baryon and meson share no common quarks ({\it e. g.} $J/\psi$ production).  

At lower energies, $t$-channel Regge trajectories associated with meson exchange can dominate.  These Regge trajectories can accommodate a wider range of quantum numbers, including charge exchange.  These trajectories are necessary to explain $\pi^+$ and $\pi^0$ production: the $\pi^+$ can be produced via trajectories that include charged pions while the $\pi^0$ can be produced via Regge trajectories that include $\omega$ mesons.   In pQCD, meson-exchange trajectories are composed primarily of quarks.  In $u$-channel production, baryon number exchange can be incorporated in pQCD by the exchange of a diquark 
accompanied by the creation of an additional $q\overline q$ pair to conserve baryon number.  

\begin{figure}[!h]
\begin{center}
    
\includegraphics[width = \columnwidth]{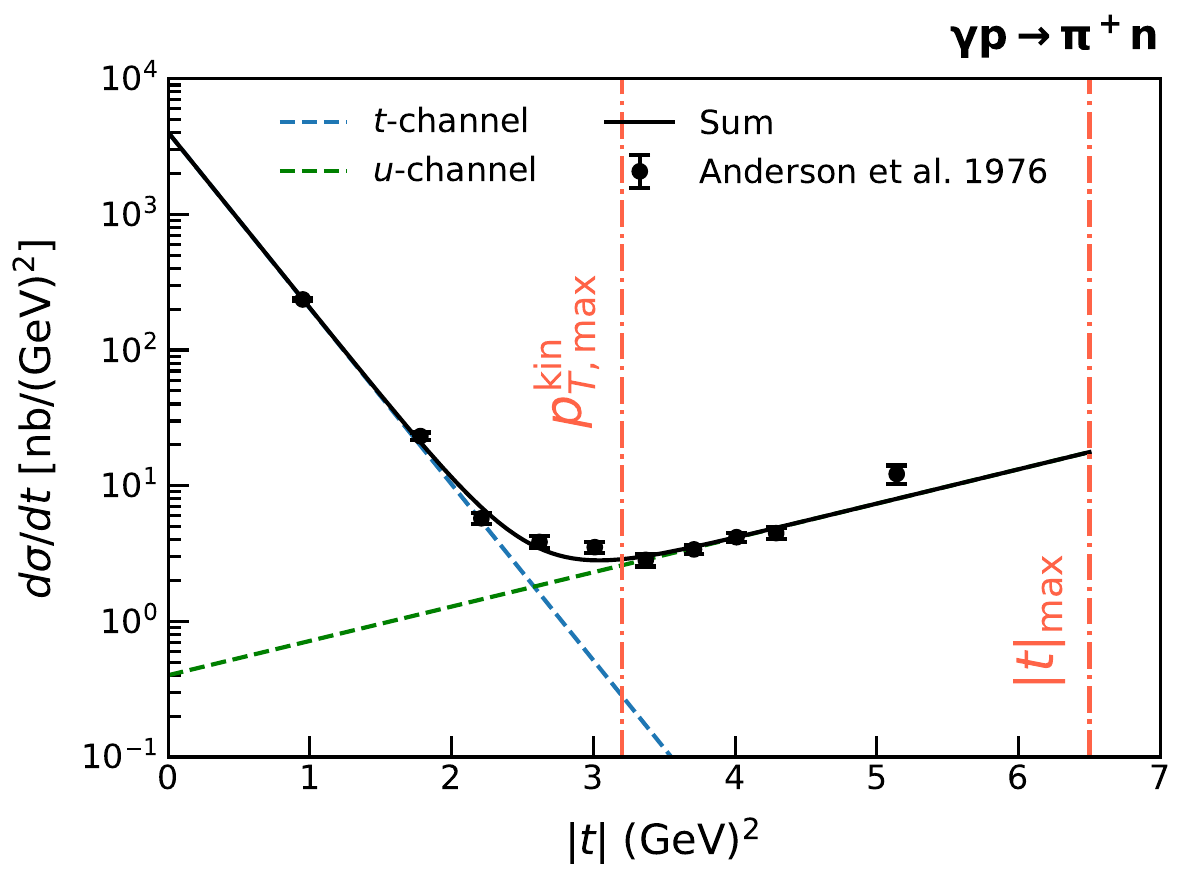}
\caption{Diffractive $\pi^+$ photoproduction off a proton. The green and blue dashed lines are the $u$ and $t$-channel contributions respectively. The solid black line is the sum of the $t$ and $u$-channel fits to the data. The different slopes of the $u$ and $t$-channel processes are obvious.  The red vertical dash-dotted lines represent the maximum kinematically allowed values of $|t|$ and $|t|_{\rm max}$, on the right with maximum $p_T$ and $p^{\rm kin}_{T\,{\rm max}}$ near the center of the plot. Adapted from Ref. \cite{anderson1976measurements}.}
\label{fig:TandU}
\end{center}
\end{figure}

\begin{figure*}[t]
\begin{center}

\includegraphics[width=0.95\textwidth]{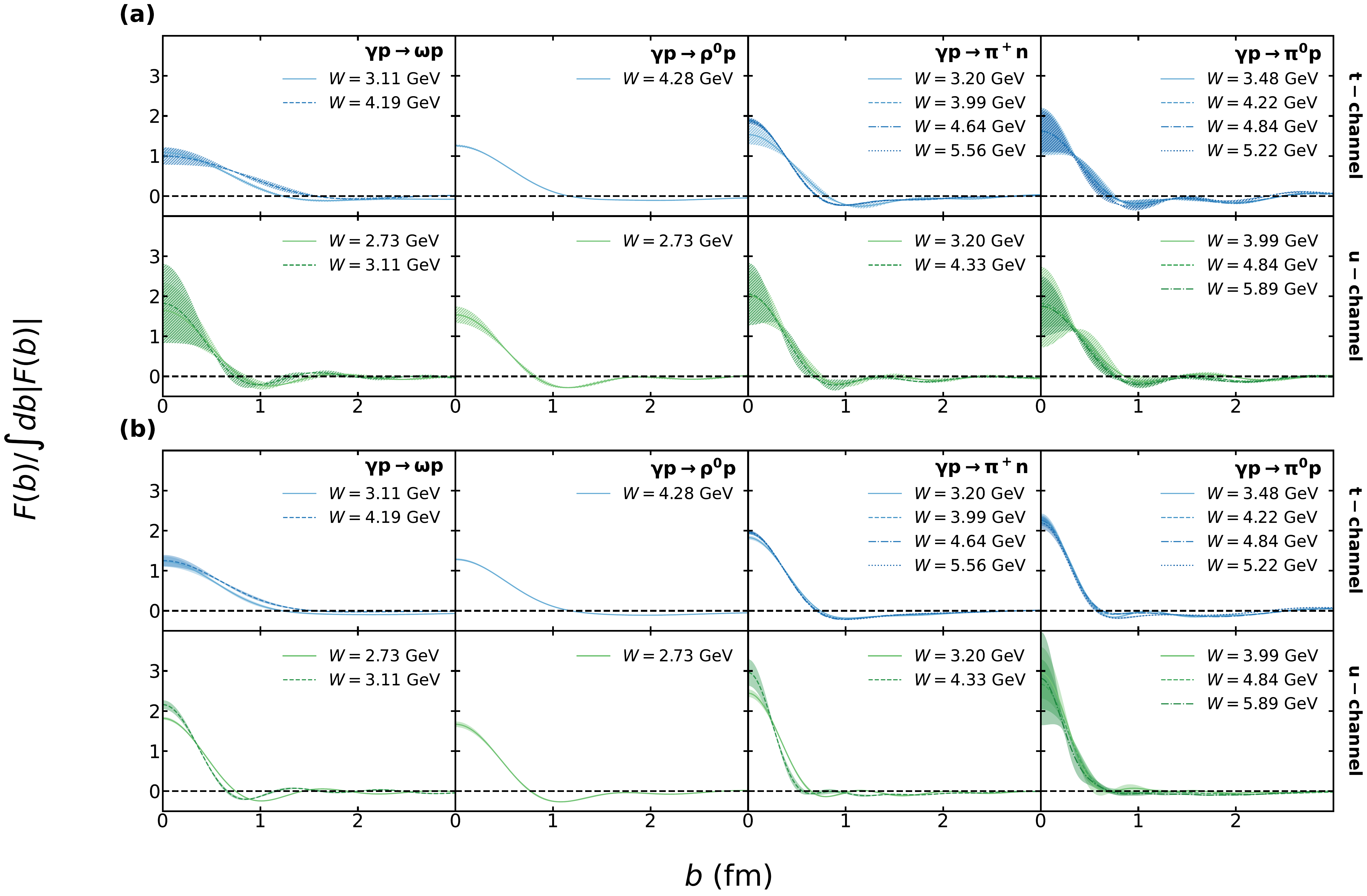}
\caption{Normalized $F(b)$ distributions obtained via Fourier-Bessel transforms, as discussed in the supplemental material, for the two different finite-window control methods. In (a) no extrapolation is used, only an incompleteness uncertainty. The incompleteness uncertainty and statistical uncertainty are added in quadrature and drawn as shaded bands around the central curves. The results in (b) use an extrapolation to $p^{\rm kin}_{T, \, \rm max}$. Minor finite window effects still exist in (b) since the integral is formally defined from $p_T = 0$ to $p_T = \infty$, and $p^{\rm kin}_{T, \, \rm max}$ is finite.}
\label{fig:Fofb}
\end{center}
\end{figure*}

Figure~\ref{fig:TandU} shows a measurement of the exclusive photoproduction cross section of $\pi^+$ production from a proton target for $W = 2.90$~GeV \cite{anderson1976measurements}. The data are measured over a large range of $t$. Since $t + u$ is a constant, larger $t$ corresponds to smaller $u$. The data shown in Fig.~\ref{fig:TandU} can be described by a sum of exponentials for $X = u$ and $t$, indicated by the dashed lines, as in  Eq.~(\ref{eq:regge}). The kinematic maximum of $t$, $t_{\rm max}$, (or, equivalently, of $u$) is indicated by the vertical dot-dashed line on the right-hand side. Also shown is the maximum allowed $p_T$ of the produced meson, $p^{\rm kin}_{T\,{\rm max}}$, separating the $t$ and $u$-channels.  The slope of the $u$-channel process is larger than for the $t$-channel equivalent. $t$-channel cross section decreases more rapidly than the $u$-channel cross section changes with $t$. The  normalization factor is also different in the two cases, with $A_t > A_u$. 
Since impact parameter, $b$, is conjugate to $p_T$, it is possible to Fourier-Bessel transform the $t$- and $u$-channel amplitudes of exclusive meson photoproduction reactions to determine the spatial distribution, $F(b)$,  of the scattering source. Details of the method are given 
in the supplemental material.  Since the transformed data are quasi-exponential, the resulting impact parameter distributions are quasi-Gaussian.
We use these impact parameter distributions to extract
the half-width at half-maximum (HWHM) of $F(b)$, which is taken  to be a measure of the transverse source size.  The data studied here, from a variety of experiments, are discussed in the supplemental material.
\begin{figure}[!h]
\begin{center}
\includegraphics[width=0.9\columnwidth]{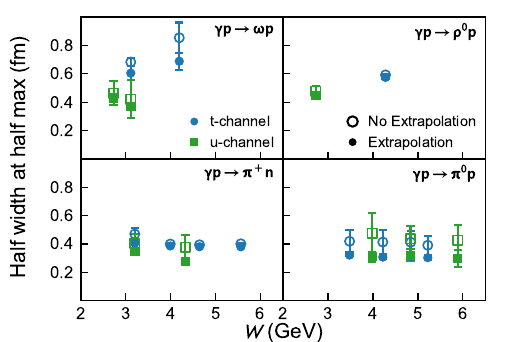}
\caption{The HWHM of the $F(b)$ distributions. Blue circles display the $t$-channel results while the green squares denote those for the $u$-channel. Open markers are the HWHM with no extrapolation used in the Fourier transform.  The solid markers indicate the results using extrapolations. }
\label{fig:hwhm}
\end{center}
\end{figure}
Because the data cover a finite range in $u$ and $t$ while Fourier-Bessel transforms are defined from $p_T = 0$ to $\infty$, it is necessary to account for finite range effects (`windowing') {it on} the uncertainty. Two 
methods were developed to account for windowing:
`extrapolation', and `no extrapolation'. 
\begin{table*}[!t]
\setlength{\tabcolsep}{10pt}
\renewcommand{\arraystretch}{1}
\begin{center}
\begin{tabular}{ l l c }
\hline
\hline
Reaction & Sensitive to & RMS Radius (fm) \\ 
\hline
Low-energy forward VM photoproduction & High-$x$ quarks and gluons & $0.67-0.77$  \\ 
Forward $\pi^0$ and $\pi^+$ photoproduction & High-$x$ quarks & $0.33-0.45$ \\ 
Backward meson production & Baryon number & $0.33-0.53$ \\
\hline
High-energy forward VM photoproduction & Gluons & $0.86-0.99$  \\
Electron scattering & Net charge & $0.69-0.72$   \\
Hydrogen spectroscopy & Charge & 0.69 \\ 
\hline
\hline
\end{tabular}
\caption{Different methods of determining 2D radii, the characteristics of the proton they are sensitive to, and their inferred 2D radii. The top three values are extracted from this work while, below the line, results from other measurements are shown. `VM' stands for vector meson. See text for discussion.}  \label{tab:radi}
\end{center}

\end{table*}
The cross sections are  transformed according to Eq.~(\ref{Fs}). Both 
methods discussed in the supplemental material are applied to each dataset. The resulting $F(b)$ distributions are shown in Fig.~\ref{fig:Fofb} (a) with no extrapolation and in (b) with extrapolation. The upper panels show the $t$-channel distributions while the lower panels display $u$-channel distributions. Without extrapolation, a substantial incompleteness error is visible for small values of $b$ since $p_T \propto 1/b$. The $u$-channel datasets have a larger incompleteness error because their cross sections fall more slowly in the unmeasured regions then the $t$-channel cross section. The extrapolation technique reduces bias 
due to windowing. 

Based on the $F(b)$ distributions, the HWHM is calculated and displayed as a function of $W$ in Fig.~\ref{fig:hwhm}. The HWHM was chosen since it is robust against the oscillations around $F(b) = 0$ at large $b$ that arise from finite window effects. The HWHMs are given both with and without extrapolation; the HWHMs with extrapolation tend to be smaller than their non-extrapolated counterparts, by an average of 14\% for $t$-channel scattering and 20\% for $u$-channel reactions.  The non-extrapolated radii are biased because of the partial coverage in $p_T$.  Because the extrapolated results should be much less biased, we use them in our discussion. A detailed tabulation of the HWHM and the covereage in $p_T$ space is provided in the supplementary material.

In the $\pi^0$ case, both channels are consistent within uncertainties with both methods. However, the remaining three systems show a smaller HWHM for the $u$-channel data relative to the $t$-channel, regardless of the 
method used with a HWHM around 0.4~fm. This conclusion strengthens greatly when extrapolation is included. 

% Table \ref{tab:summary_hwhm} characterizes the $F(b)$ distributions by their HWHMs. The HWHMs are given both with and without extrapolation; the HWHMs with extrapolation tend to be smaller than their non-extrapolated counterparts, by an average of 14\% for $t$-channel scattering and 20\% for $u$-channel reactions.  The non-extrapolated radii are biased because of the partial coverage in $p_T$.  Because the extrapolated results should be much less biased, we use them in our discussion.

% ML Commented above line that references fig 1, which he has removed because of legnth. The statement starting with " The HWHMs are given both with and without extrapolation ..." is moved to to the paragraph above that describes fig 3.

The HWHMs for all of the $u$-channel transforms are in the range $0.28-0.45$~fm while the $t$-channel HWHMs cover a wider range, $0.30-0.69$~fm, with a trend toward smaller sizes as $W$ increases. The $t$-channel HWHMs show clear differences depending on the final-state meson.  The $\omega$ and $\rho^0$ production channels that can proceed via Pomeron exchange (i.~e.\ gluons) have HWHMs of 0.57 to 0.69~fm, while the $\pi^+$ and $\pi^0$, which are likely produced via meson or quark exchange, have HWHMs of 0.30 to 0.40~fm - significantly smaller.  This is consistent with another measurement which found that, in a proton, quarks with $x \gtrsim 0.3$ have  HWHMs of around 0.4~fm~\cite{Miller:2008jc}, comparable to the HWHMs of the $\pi^+$ and $\pi^0$ measurements.
All of the measurements used here were taken at 
similar photon energies, allowing for more direct comparisons than higher energy photoproduction measurements.  One drawback of these lower energies, however, is that the $\omega$ and $\rho^0$ are also produced through Reggeon exchange; these Reggeons are expected to be mostly quarks.

Table \ref{tab:radi} puts these numbers in perspective, comparing them with proton radii measured using other techniques which are sensitive to different aspects of the proton, as given in the second column.  The third column gives the 2-dimensional (2D) RMS radii, $r_2$. The HWHMs of the results presented here are multiplied by 1.1775, appropriate for a Gaussian distribution, to convert from HWHM to RMS.  Many of the other results are given as 3-dimensional (3D) radii, $r_3$. They are converted to $r_2$ following
~\cite{Strikman:2003gz,Caldwell:2010zza},
\begin{equation}
\langle r_{2}^2\rangle = \frac{2}{3}\langle r_{3}^2\rangle \,.
\end{equation}

As previously discussed, 2D radii of protons have also been obtained
using high-energy vector meson photoproduction.  
In these processes, the photons fluctuate to quark-antiquark dipoles which scatter elastically from the target nucleus.  This colorless hadronic scattering is expected to be Pomeron mediated, thus primarily probing the gluon distribution of the target.

One recent H1 $\rho^0$ production study from HERA \cite{H1:2020lzc}, at center of mass energies of 20-80 GeV, found an exponential slope, $b_\rho=9.61 \pm 0.15 \, ({\rm exp.}) ^{+0.15}_{-0.20} \, ({\rm syst.})$~GeV$^{-2}$, equivalent to a radius of 1.24 fm.  
Using this value, we can subtract the square of the $\rho^0$ radius, $r_\rho^2=0.54\pm 0.02$~fm$^2$, by quadrature, obtaining $r_p = 0.99$~fm, albeit with considerable uncertainty. A ZEUS measurement of $J/\psi$ photoproduction found $b_{J/\psi}=4.6$ GeV$^{-2}$, \cite{ZEUS:2002wfj}, giving $r=0.86$~fm, before subtraction of the unknown $J/\psi$ radius.  
These measurements, at much higher energy than the $\omega$ and $\rho$ data used in our analysis, probe gluons at much smaller $x$ values, so the radii should be larger.

A similar meson-size correction should apply to the lower-energy data used here. Because the size of these corrections is difficult to estimate, %accurately, 
an alternate approach is to take the quoted sizes as upper limits because any meson-size correction will reduce the RMS radii.   Meson-size effects should be comparable in the production of mesons with similar masses at comparable photon energies. 

Other experiments have measured the proton charge radius more directly using electron scattering.  This approach does not suffer from the meson-size uncertainty.  
Some experiments find $r_3 \approx 0.84$~fm while others find $r_3 \approx 0.88$~fm \cite{Gao:2021sml}, with considerably smaller uncertainties than the difference between them.  The resulting 2D radii are similar to the gluon radii inferred from our lower-energy data. 

Alternately, one can measure the charge radius using low-energy experiments, such as measurements of the muonic radius of hydrogen, giving a charge radius of $\langle r_3^2\rangle^{1/2}\sim 0.84$~fm. Measurements of the hydrogen Lamb-shift spectra  ~\cite{Bezginov:2019mdi} also find similar  values of $\langle r_3^2\rangle^{1/2}$.

Table \ref{tab:radi} shows a clear distinction between measurements of the charge or mass (gluon) distribution with radii larger than 0.67~fm, while measurements sensitive to net quark or baryon number find smaller values, $0.33 - 0.53$~fm. %Table \ref{tab:radi} shows a clear grouping.  Measurements of the charge or mass (gluon) distribution find radii larger than 0.67~fm, while measurements that should be sensitive to net quarks or baryon number find smaller values, 0.33 to 0.53~fm. 
Measurements sensitive to the charge or mass involve diverse techniques, including low-energy atomic physics probes such as the energy levels of muonic hydrogen, elastic electron-proton scattering, and vector meson photoproduction. All of the charge-radius measurements find similar radii, around 0.7 fm to 'around 0.67-0.72 fm.  The gluonic radii are still larger, 0.86-0.99 fm.  Those sensitive to baryon number, including backward production and photoproduction reactions coupling to net quarks have considerably smaller radii.  Thus, 
baryon number is confined to a noticeably smaller region than the baryon as a whole.

We have measured the effective sizes (HWHMs) of photoproduction targets, responsible for backward photoproduction, and compared them to the size of the targets in $t$-channel reactions at comparable energies as well as other measurements of the size of the proton.  Reactions involving backward production find effective sizes that are generally smaller than those involving forward production, and smaller than measurements of the proton charge radius.  This shows that baryon number is likely confined to the central region of a proton.

Looking ahead, backward production data from a wider range of mesons would help differentiate between models.  Heavier mesons such as the $\phi (s \overline s)$ and $J/\psi (c \overline c)$ 
which share no valence quarks with the target protons are of particular interest.  They should be more suppressed in models involving diquark exchange than in models involving baryon junction exchange. There is already a hint of near threshold $u$-channel $J/\psi$ photoproduction~\cite{GlueX:2023pev}.  

The future Electron-Ion Collider \cite{AbdulKhalek:2021gbh} is well suited for studying backward production. The rates for many mesons are high enough to permit detailed kinematics studies \cite{Cebra:2022avc}.  Backward virtual Compton scattering is of particular interest because the photon could directly couple to the baryon.  
eSTARlight simulations \cite{Lomnitz:2018juf}  of backward virtual Compton scattering and $\pi^0$ production
predict that these two reactions occur at high rates and, crucially, can be separated from each other \cite{Sweger:2023bmx}.  The $Q^2$ dependence of the cross section is also sensitive to the production mechanism.

\textit{Acknowledgements} This work is supported in part by the U.S. Department of Energy, Office of Science, Office of Nuclear Physics, under contract numbers DE-AC02-05CH11231 (SRK), DE-SC0026252 (GAM) and DE-AC52-07NA2734 (RV).  This material is also based upon work supported by the National Science Foundation under Grant No. 250721 (ML, ZS). GAM thanks the UC Berkeley Physics Department NSF cooperative agreement 2020275, and  Lawrence Berkeley Lab, U.S. Department of Energy, Office of Science, Office of Nuclear Physics, under Contract No. DE-AC02-05CH11231  for their hospitality during the initial stages of this work.
\bibliographystyle{apsrev4-1} 
\bibliography{inco}
%\paragraph*{Author contributions:}
%All authors contributed to writing this paper.  SRK conceived the idea and framed the problem.  ZS did the initial calculations, while MCL did the final calculations that are used here, including making the plots and tables.  RV and GAM contributed to the theoretical formulation of the Fourier transformation and interpretation of the results.
%\paragraph*{Competing interests:}
%There are no competing interests to declare.
%\paragraph*{Data and materials availability:}
%All data are available in the manuscript or the supplemental materials.

%%%%%%%%%%%%%%%% SUPPLEMENT LIST %%%%%%%%%%%%%%%

% List the contents of your supplemental Materials, including the numbers of any
% supplemental figures, tables, external data files etc. and any references that are
% cited only in the supplement. In this example, refs. 7-8 are cited only in the supplement.
% Fill out your numbers accordingly and delete any lines that aren't applicable.

\section*{supplemental Materials}
\subsection*{Methods}
Figure~\ref{fig:diagrams}(b) shows that $u$ plays the same role for backward production that $t$ does for forward production.   Although the standard two-body cross section, $d\sigma/dt$ is expressed as proportional to the square of the invariant amplitude (see {\it e.g.} Eq.~(49.33) of Ref.~\cite{ParticleDataGroup:2024cfk}), $d\sigma/du$ is also proportional to the square of the invariant amplitude.  The cross section may be written more generally as
\begin{equation}
    {d\sigma\over dX}={1\over 64 \pi s}{1\over |{\bf p}_{\rm 1cm}|^2}|{\cal M}|^2 \, \, ,
\label{basic}
\end{equation}
where $\cal M$ is the invariant amplitude, an evaluation of relevant Feynman diagrams. The factors multiplying $|{\cal M}|^2$ arise from incident flux and phase space factors.  The values of $\cal M$ depend on the kinematics of the measurements where either $|u|$ or $|t|$ is small.   The goal of this work  is to understand the spatial structure of the amplitude, $\cal M$. 

In the case of forward processes (small $|t|$), calculations of  $\cal M$ depend on the momentum transfer ${\bf p_\gamma}-{\bf p_V}$. In the high energy limit of the photon,  the energies of the photon and vector meson are approximately equal, resulting in $t\approx-({\bf p_\gamma}-{\bf p_V})^2=-p_T^2$. 
Thus $\cal M$ depends on $p_T$ (and $s$). On the other hand, in backward processes the energies of the photon and proton are approximately equal with
$u\approx-({\bf p_\gamma}-{\bf p_p})^2=-p_T^2$, ~\cite{Perl1974High}.  
In either case, the longitudinal momentum transfer can be neglected, as in eikonal approximations~\cite{Glauber:1970jm,Sakurai:2011zz} of the cross section. Spatial distributions corresponding to amplitudes measured in momentum-transfer processes are obtained by making Fourier transformations ~\cite{Hofstadter:1956qs}. A two-dimensional Fourier transform is sufficient here because the scattering amplitude depends only on ${\bf p}_T$. Thus, we obtain
\begin{eqnarray}
    \widetilde {\cal M}({\bf b}) & = & \int {d^2p_T\over (2\pi)^2}e^{-i {\bf p_T}\cdot{\bf b}}{\cal M}({\bf p}_T).
\end{eqnarray}
Because no spin dependence is measured, the scattering amplitude effectively depends only on the magnitude of ${\bf p}_T$. The angular integration of the exponential factor then leads to the result 
\begin{eqnarray}  \widetilde {\cal M}({ b}) &  = &{1\over  2\pi} \int_0^\infty p_T dp_T J_0(bp_T) {\cal M}(p_T),
\end{eqnarray}
where $J_0$ is the cylindrical Bessel function. 
To proceed, we need to relate $\cal M$ to  the measured cross sections, $d\sigma/dX$, Eq.~(\ref{basic}).
At small values of $|t|$, 
$\cal M$ is  primarily imaginary~\cite{horn1973}, while for small values of $u$ the amplitude is mainly real~\cite{Clifft:1977yi}.  In either case, we obtain ${\cal M} =\sqrt{64\pi s }\;{\bf p}_{\rm 1cm} \sqrt{d\sigma/dX}$. 
The net result is then
\begin{eqnarray}
 F(b)=\int_0^\infty p_T dp_T J_0(bp_T) \sqrt{d\sigma\over d X} \, \,  ,
 \label{Fs}
    \end{eqnarray}
where the kinematic factor $\sqrt{64\pi s }\;{\bf p}_{\rm 1cm}/(2\pi) $ is ignored because it cancels in the ratio we analyze, 
 $F(b)/\int db|F(b)|$. 
Equation~(\ref{Fs}) thus applies to processes in which either $|u|$ or $|t|$ is small.
It is necessary to relate $p_T$ to either $t$ or $u$, depending on the kinematics.

In Eq.~(\ref{Fs}),  the use of a two-dimensional Fourier transform is  compatible with standard treatments necessary for nucleons because they contain  particles (partons) that move with relativistic speeds close to that of the speed of light.   Three-dimensional  transforms work for nonrelativistic structure but fail when relativistic effects become significant \cite{Gao:2021sml,Miller:2010nz,Miller:2025zte}.  Similar techniques have been used to determine the size and shape of nucleons and nuclei \cite{Alvensleben:1970uw,Klein:2019qfb}.  Measurements of $d\sigma/dt$ have been used to determine the transverse target profile of protons and heavy nuclei \cite{STAR:2017enh} using Eq.~(\ref{Fs}) with $X = t$.
There are multiple caveats associated with this method \cite{Klein:2021mgd}. 

Some analyses have fit the cross section to an exponential, as in Eq.~(\ref{eq:regge}), with the slope related to the target size. At high energies, where $X_{\rm min} \approx 0$, more complex functional forms, such as $d\sigma/dt \propto \exp(-D|t| + Ct^2)$ have also been used.  The slope, $D$, of the exponential in Eq.~(\ref{eq:regge}) strongly depends on the mass of the produced vector meson in $t$-channel tomography.  The extracted value of $D$ decreases systematically with increasing vector meson mass, from 9.61~GeV$^{-2}$ for $\rho^0$ photoproduction to $\sim 4-5$~GeV$^{-2}$ for the $J/\psi$ \cite{Crittenden:1997yz,H1:2013okq}.  This change in slope can be attributed to fluctuations of the virtual photon into a $q \overline q$ dipole, with finite transverse size, that interacts with the target to produce the vector meson.  Thus the width of $F(b)$ determined from Eq.~(\ref{Fs}) is a convolution of the transverse size of the dipole and the transverse size of the target. The change in $D$ with vector meson mass is associated with a narrowing of the peak of $d\sigma/dt$.  With Pomeron exchange, $D$ also decreases slowly with increasing collision energy, an effect known as diffractive peak shrinkage.  Shrinkage can be explained by an increase in target size as partons with lower and lower momentum fractions, $x$, are involved in the reaction.  We avoid bias due to peak shrinkage by comparing forward and backward reactions at similar photon energies. 

\begin{figure*}[t]
\begin{center}
\includegraphics[width=0.95\textwidth]{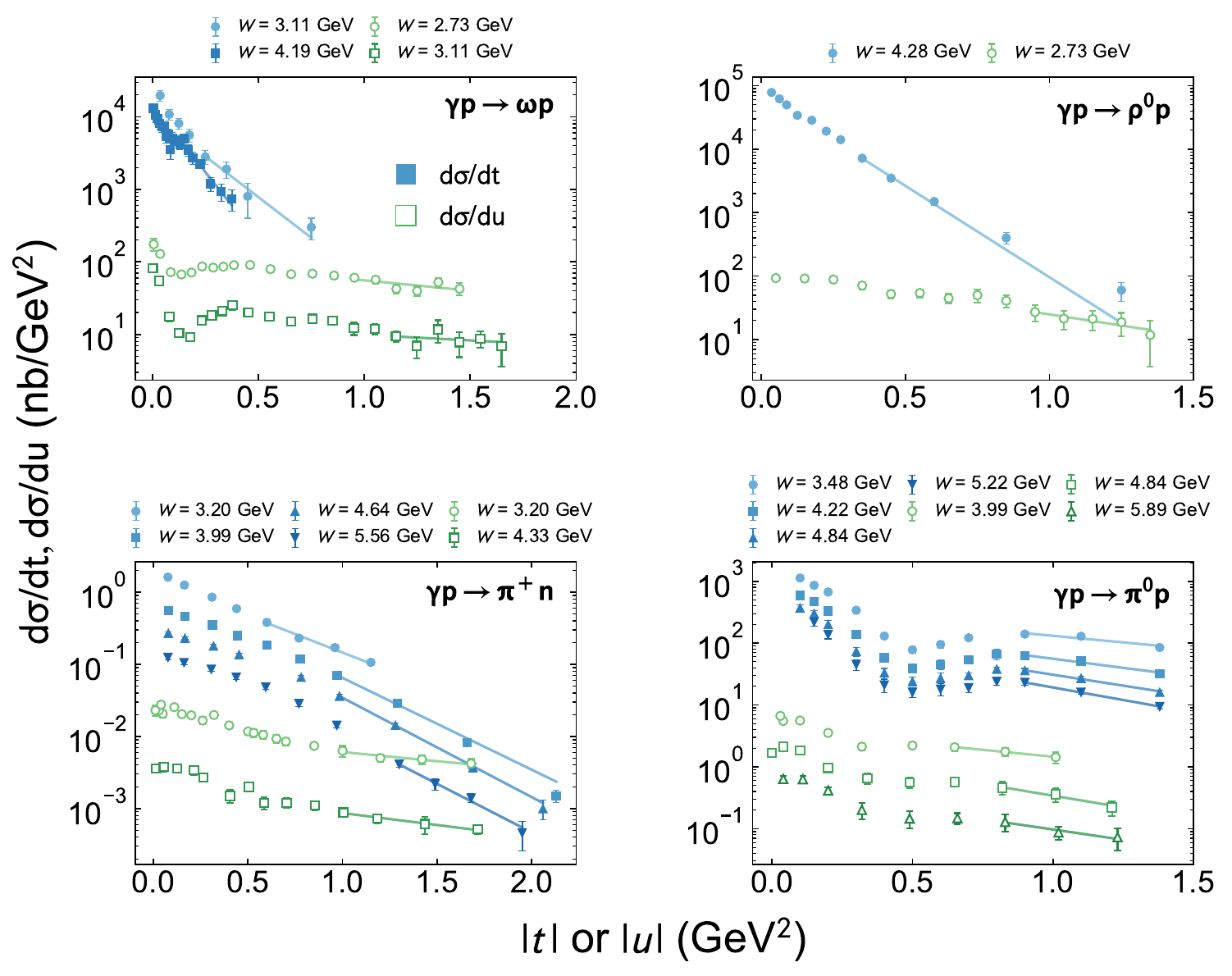}
\caption{Cross section measurements for meson photoproduction off proton targets. The hollow green points correspond to $u$-channel processes. The solid blue points represent $t$-channel reactions. Exponential function fits to the tails of the distribution, indicated by the solid lines, are used to estimate the incompleteness of the integration. The number of points used in the fit, chosen to best represent the shape of the tail, varies with each dataset.}
\label{fig:cross_sections}
\end{center}
\end{figure*}

This same formulation can be applied to backward production where $\cal M$ is evaluated as a function of  $u$ in Eq.~(\ref{Fs}).  In this case, the imaging probes the distribution of baryon number in the target.

The finite range of $p_T$ in the data introduces some potential bias, known as the incompleteness uncertainty.  Fourier transforms cover the full range of $p_T$, from zero to $\infty$, while the data is cut off at a finite value of $p_T$. In reality, four-momentum conservation imposes a maximum kinematically allowed $p_T$, 
$p^{\rm kin}_{T, \,{\rm max}}$, above which the cross section must be zero.   The data we analyze does not reach this kinematic limit.  The lower $p_T$ cutoff, equivalent to convoluting the data with a Heaviside function, introduces a windowing artifact in the transform unless the maximum $p_T$ is sufficiently high for $d\sigma/dX$ to be negligible. 
In the case of heavy-ion targets, it is necessary to integrate up to the third diffractive minimum to avoid significant artifacts \cite{Aschenauer:2025mku}.

The incompleteness uncertainty has been studied for the electromagnetic form factors of the proton~\cite{Venkat:2010by}.  That study related the maximum value of $Q$ used in the Fourier-Bessel transform to the uncertainties on the the extracted transverse charge and magnetization densities. 

Here, two techniques are used to account for the finite window effects. The first treats the unmeasured region as a contributor to the systematic uncertainty in the measurement, as motivated by Ref.~\cite{Venkat:2010by}. A given dataset is first transformed according to Eq.~(\ref{Fs}) using the maximum $p_T$ of the data ($p^{\rm data}_{T, \, \rm max}$) 
as the upper bound of the integral. Then, an ``incompleteness error" is calculated. This is done by fitting the last 3-5 points of a given dataset with an exponential function and then extrapolating the fit to $p^{\rm kin}_{T,{\, \rm max}}$. 
The number of points used to fit the exponential function to the data are chosen to minimize statistical fluctuations while still capturing the shape of the tail at high $t$ or $u$. The incompleteness error, 
\begin{equation}
    \Delta_{\rm inc} =  \left| \frac{1}{2\pi} \int_{p^{\rm data}_{T, \, \rm max}}^{p^{\rm kin}_{T, \, \rm max}} p_T dp_T J_0(bp_T) \sqrt{\frac{d\sigma}{dX}}\right| \ , 
\end{equation}
is an estimate of the contribution to the integral missed due to an incomplete measurement of the kinematically allowed $p_T$ range. The closer 
the $p^{\rm data}_{T, \, \rm max}$ is to $p^{\rm kin}_{T,{\, \rm max}}$, the smaller the incompleteness error. Likewise, a larger extrapolated exponential slope will result in a smaller incompleteness error. The incompleteness error is treated as a systematic error.

The second method attempts to correct for systematic effects. Instead of evaluating an incompleteness error, an extrapolation from 
the maximum measured $p_T$ to $p^{\rm kin}_{T, \, {\rm max}}$ is 
added to the measured data. In this way, the integral is calculated over the entire kinematically allowed $p_T$ range. This introduces some bias into the procedure and the choice of model used in the extrapolation could impact the extracted $F(b)$. However, the use of exponential functions is standard in such cases due to their success in describing various diffractive processes.  We present results with and without this extrapolation, including the incompleteness error in the uncertainty for the no-extrapolation results. 

The data sets we study typically have on the order of a dozen or fewer measurements of $d\sigma/du$ for a given $W$. 
Therefore, we perform a finite Fourier transform, using these points and linearly interpolating between adjacent points. Because the spectra are quasi-exponential, logarithmic interpolation was also tried.  
However, this did not significantly change $F(b)$. 

The uncertainties are estimated by resampling.  For each transform, we generate 1000 pseudodata sets, with each new set generated randomly using the data points and their Gaussian errors.   Each pseudodata set is then Fourier transformed, building up a statistical error band and making calculations of uncertainties on the widths of the distribution possible.

\subsection*{Description of the datasets used}
As outlined in the main text, we study backward photoproduction of $\pi^0$, $\pi^+$ (with a final-state neutron), $\rho^0$, and $\omega$, which are then compared to $t$-channel data at similar collision energies. We chose these measurements because of the relatively abundant data, including at multiple photon-proton center-of-mass energies, $W$.  We select data with $W > 2.5$~GeV to avoid the baryon resonance region. Figure~\ref{fig:cross_sections} shows the cross sections studied for the four reactions discussed here, described in more detail below. 
Best fit lines, drawn for each data set, are used to calculate the incompleteness error.  They are also used in the extrapolation technique. 

The backward $\pi^0$ data \cite{Tompkins:1969jd} was taken using a tagged photon beam at the Stanford Linear Accelerator Center (SLAC).  The reaction products were observed in a large (for the time) spectrometer, with hodoscopes for tracking in a magnetic field and Cherenkov counters for particle identification.  The data covers the photon energy range from 3.5 to 5.9~GeV. Because the 3.5~GeV data covers a rather limited $u$ range, it was not used here. There is a possible diffractive dip. The forward $\pi^0$ data \cite{anderson1971high}, using the same experimental apparatus as the corresponding $u$-channel measurement, and is measured at photon energies of 6, 9, 12, and 15~GeV. A clear diffractive peak is seen across all forward $\pi^0$ data sets around $|t| = 0.5$~GeV.

The $u$-channel $\pi^+$ data also came from SLAC \cite{piplusback}, using photon beams from 4.1 to 14.8~GeV, with a spectrometer for pion identification. A peak is visible close to $|u| = 0$.  No diffractive dip is observed. Only the 5.0 and 9.5~GeV photon data are used, chosen for their large coverage in $u$. Forward $\pi^+$ cross sections were also measured with the same SLAC spectrometer and hodoscope setup \cite{piplusdatatchannel}. In this case, the data were measured down to very small values of $t$. There is a sharp peak close to $t = 0$, associated with single pion exchange.  It was removed from the present analysis which only includes points with $|t| > 0.07$~GeV$^2$.

Backwards $\omega$ production data comes from the Daresbury Laboratory electron synchrotron NINA \cite{Clifft:1977yi}, which produced tagged photons between 2.8 and 4.8~GeV. Final-state protons 
were tagged using a rotating spectrometer arm while Cherenkov counters were used to isolate pions from $\omega$ decays. The $\omega$ data exhibits a clear diffractive dip around $u = 0.2$~GeV$^2$. The $t$-channel $\omega$ data were measured in two experiments \cite{Abramson:1976ks, Ballam:1972eq}. Reference~\cite{Abramson:1976ks} used a tagged photon beam and a missing mass technique to find the outgoing proton. Reference~\cite{Ballam:1972eq} used a hydrogen bubble chamber exposed to linearly polarized photons.  This experiment also measured forward $\rho^0$ data. Backward $\rho^0$ data, also taken at NINA \cite{Clifft:1976be}, displays no diffractive minima.

There is one clear distinguishing characteristic difference among the various spectra: while some of the distributions could be well described by an exponential, others exhibit clear diffractive minima. A similar observation holds for backward production using pion beams: some reactions exhibit diffractive minima while others do not \cite{Berger:1971zz}. Diffractive minima are associated with destructive interference effects, as well as with relatively sharp density drops at the edge of the target. 

Backward production of the $\omega$, for example, shows a clear diffractive minima, while forward $\omega$ production appears more exponential.  The $\pi^+$ backward production data show some structure while forward production could easily be fit by an exponential.  While there is no strict diffractive minimum for the $\pi^0$, the data exhibits a multi-slope characteristic, with a relatively hard tail extending to large $|u|$.  The forward production data is similar, although it has a somewhat more apparent diffractive peak.  

These shape differences may be associated with the type of quasiparticle being exchanged. Because Pomerons are expected to be mostly gluonic, they are sensitive to the gluon distribution in the target.  The Bjorken $x$ range depends on the photon energy, decreasing as the energy rises.  These data indicate that the gluon distribution is smooth, without a hard edge. The $\pi^+$  may arise from pion exchange while the $\pi^0$ could be produced via $\omega$ exchange. 
\vspace{2 cm}

\subsection*{The relationship between $p_T$, $t$, and $u$}
The transformations presented in this article are defined over the transverse momentum of the photoproduced meson. However, the cross sections are reported as differential measurements in $t$ or $u$.  Thus, it is necessary to parameterize $d\sigma/dt$ and $d\sigma/du$ in $p_T$. 

There are exact expressions that relate $t$ and $u$ to $p_T$.  First, one must define $t$ and $u$ in terms of $\theta$, the angle between the incoming photon and the meson produced in the center of mass frame. In the $t$-channel
\begin{equation}
\cos\theta \;=\; -\,\frac{\,G \;-\; 2 W^2\!\left( 1 + \tfrac{t+Q^2}{W^2 - m_p^2}\right)}{\sqrt{\,G^2 - 4 W^2 m_p^2\,}} ,
\end{equation}
where $m_p$ and $m_M$ are the proton and meson masses respectively, and $G = m_p^2 + Q^2 + W^2.$  Here $Q$ is the momentum transfer and $W$ is the center of mass energy.  A similar expression can be found for the $u$-channel 
\begin{equation}
\cos\theta \;=\; -\,\frac{\,2 W^2 G_0 \!+\! G_1 \left(W^2 + m_M^2 - m_p^2\right)}{\sqrt{\;\big(G_1^2 - 4 W^2 m_p^2\big)\,\big(G_2^2 - 4 W^2 m_p^2\big)\;}} ,
\end{equation}
where  $G_0 = u-m_p^2-m_M^2$, $G_1 = W^2 + m_p^2 + Q^2$ and $G_2 = W^2 - m_M^2 + m_p^2$. 

The transverse momentum of the produced meson can then be expressed as $p_T = p^{\rm kin}_{T, \,{\rm max}} \,  \sin \theta$, where
\begin{equation}
p^{\rm kin}_{T, \,{\rm max}} \;=\; \frac{\sqrt{\lambda\!\left(W^2,\, m_p^2,\, m_M^2\right)}}{2W} \, \, 
\end{equation}
where $\lambda(a,b,c) \;=\; a^2 + b^2 + c^2 \;-\; 2\,(ab + bc + ca) $ is the K\"allen function. In our case $Q^2 = 0$.  The cross sections, parametrized as a function of $p_T$ using the above prescription, are used to evaluate the integrals as a function of $p_T$.

\subsection*{Tabulation of HWHM}
Table \ref{tab:summary_hwhm} shows the HWHM results for all  datasets studied. To better understand the incompleteness uncertainty, the maximum $p_T$ reached in each dataset is also reported.
\begin{table*}[!t]
\setlength{\tabcolsep}{6.5pt}
\renewcommand{\arraystretch}{0.85}
\centering

\begin{tabular}{c c c  c c c c}
\hline\hline
Meson & Channel & $W$ (GeV) & HWHM (fm), No Extrap. & HWHM (fm),  With Extrap. & $p_{T, \, \text{max}}^{\text{data}}$ (GeV) & \%  Coverage\\
\hline
$\omega$    & $t$ & 3.11 & $0.683 \pm 0.018 \pm 0.025$ & $0.605 \pm 0.047$ & 0.76 & 58\% \\
$\omega$    & $t$ & 4.19 & $0.855 \pm 0.008 \pm 0.107$ & $0.689 \pm 0.060$ & 0.58 & 30\% \\
$\omega$    & $u$ & 2.73 & $0.464 \pm 0.002 \pm 0.085$ & $0.430 \pm 0.004$ & 0.96 & 91\% \\
$\omega$    & $u$ & 3.11 & $0.422 \pm 0.003 \pm 0.134$ & 
$0.368 \pm 0.010$ & 1.09 & 84\% \\
\hline
$\rho^0$    & $t$ & 4.28 & $0.592 \pm 0.005 \pm 0.002$ & $0.575 \pm 0.010$ & 1.03 & 53\% \\
$\rho^0$    & $u$ & 2.73 & $0.478 \pm 0.004 \pm 0.034$ & $0.448 \pm 0.009$ & 0.94 & 89\% \\
\hline
$\pi^+$     & $t$ & 3.20 & $0.470 \pm 0.002 \pm 0.045$ & $0.405 \pm 0.007$ & 1.00 & 68\% \\
$\pi^+$     & $t$ & 3.99 & $0.400 \pm 0.001 \pm 0.004$ & $0.386 \pm 0.003$ & 1.34 & 71\% \\
$\pi^+$     & $t$ & 4.63 & $0.394 \pm 0.002 \pm 0.004$ & $0.381 \pm 0.003$ & 1.36 & 61\% \\
$\pi^+$     & $t$ & 5.56 & $0.401 \pm 0.002 \pm 0.005$& $0.380 \pm 0.009$ & 1.35 & 50\% \\
$\pi^+$     & $u$ & 3.20 & $0.406 \pm 0.002 \pm 0.065$ & $0.345 \pm 0.008$ & 1.18 & 81\% \\
$\pi^+$     & $u$ & 4.33 & $0.379 \pm 0.004 \pm 0.083$ & $0.277 \pm 0.018$ & 1.25 & 61\% \\
\hline
$\pi^0$     & $t$ & 3.48 & $0.419 \pm 0.002 \pm 0.079$ & $0.321 \pm 0.005$ & 1.09 & 68\% \\
$\pi^0$     & $t$ & 4.22 & $0.413 \pm 0.002 \pm 0.085$ & $0.307 \pm 0.010$ & 1.12 & 56\% \\
$\pi^0$     & $t$ & 4.84 & $0.412 \pm 0.003 \pm 0.081$ & $0.310 \pm 0.012$ & 1.14 & 49\% \\
$\pi^0$     & $t$ & 5.22 & $0.391 \pm 0.003 \pm 0.069$ & $0.304 \pm 0.014$ & 1.14 & 45\% \\
$\pi^0$     & $u$ & 3.99 & $0.476 \pm 0.004 \pm 0.145$ & $0.310 \pm 0.036$ & 0.99 & 53\% \\
$\pi^0$     & $u$ & 4.84 & $0.436 \pm 0.005 \pm 0.090$ & $0.320 \pm 0.042$ & 1.08 & 47\% \\
$\pi^0$     & $u$ & 5.89 & $0.428 \pm 0.005 \pm 0.103$ & $0.298 \pm 0.058$ & 1.10 & 38\% \\
\hline\hline
\end{tabular}
\caption{Summary of the HWHM values for $t$ and $u$-channel production at different center of mass energies $W$. Results are shown using both finite-window control methods (no extrapolation and including extrapolation).  The HWHM are reported as the central~value~$\pm$~statistical uncertainty~$\pm$~incompleteness uncertainty. The extrapolated result does not include an incompleteness error. The second to last column shows the maximum $p_T$ of the produced meson in each dataset. The final column shows the relative coverage of the data compared to $p^{\rm kin}_{T\,{\rm max}}$ in percent.}\label{tab:summary_hwhm}
\end{table*}

\end{document}